\documentclass[10pt,journal]{IEEEtran}
\usepackage{amsmath,amsfonts}
\usepackage{algorithmic}
\usepackage{algorithm}
\usepackage{array}
\usepackage{textcomp}
\usepackage{stfloats}
\usepackage{url}
\usepackage{verbatim}
\usepackage{graphicx}
\usepackage{multicol}
\usepackage{subfig}
\usepackage{cite}
\usepackage{comment}
\usepackage{multirow}
\usepackage{makecell} 
\usepackage{tabularx} 
\usepackage{amssymb}
\newcommand{\RNum}[1]{\uppercase\expandafter{\romannumeral #1\relax}}

\usepackage[dvipsnames]{xcolor}

\usepackage[colorlinks,allcolors=blue,bookmarks=false,hypertexnames=true]{hyperref} 

\begin{document}

\title{Radar Mounting Angle Estimation in Operational Driving Conditions}

\author{\IEEEauthorblockN{Simin Zhu, Satish Ravindran, Lihui Chen, Alexander Yarovoy, \IEEEmembership{Fellow, IEEE}}, Francesco Fioranelli, \IEEEmembership{Senior Member, IEEE}

\thanks{Simin Zhu, Francesco Fioranelli, Alexander Yarovoy are with the Microwave Sensing, Signals and Systems (MS3) group, Delft University of Technology, 2628 CD, Delft, The Netherlands (e-mail: s.zhu-2@tudelft.nl; f.fioranelli@tudelft.nl; a.yarovoy@tudelft.nl)\par
Satish Ravindran and Lihui Chen are with NXP Semiconductors, San Jose, CA, USA (e-mail: satish.ravindran@nxp.com).}}

\markboth{Journal of \LaTeX\ Class Files,~Vol.~14, No.~8, December~2022}%
{Shell \MakeLowercase{\textit{et al.}}: A Sample Article Using IEEEtran.cls for IEEE Journals}


\maketitle

\begin{abstract}
The problem of estimating the mounting angle of millimeter-wave automotive radars installed on moving vehicles is investigated. We address this angle estimation problem during normal driving, without relying on controlled environments, dedicated radar targets, or specially designed driving routes. To achieve this, we propose a signal processing pipeline that combines radar and inertial measurement unit (IMU) data to enable accurate and reliable estimation under realistic driving conditions. Unlike previous studies, the method employs neural networks to process sparse and noisy radar measurements, reject detections from moving objects, and estimate radar motion. In addition, a measurement model is introduced to correct IMU bias and scale factor errors. Using vehicle kinematics, the radar mounting angle is then computed from the estimated radar motion and the vehicle’s yaw rate. To benchmark performance, the proposed approach is comprehensively compared with two alternative problem formulations and four estimation techniques reported in the literature. Validation is carried out on the challenging RadarScenes dataset, covering over 79 km of real-world driving with different velocities and trajectories. Results show that stable and accurate mounting angle estimates are obtained within approximately 25 seconds of driving. To the best of our knowledge, this is the first study to demonstrate that automotive radar mounting angles can be estimated during complex, real traffic conditions using only onboard sensor data.
\end{abstract}


\section{Introduction}\label{chap_5_introduction}
\IEEEPARstart{O}{ver} the past decade, there has been increasing attention to advanced driver assistance systems (ADAS) and autonomous vehicles. To achieve safe and reliable autonomous driving, vehicles must be able to perceive complex environments with high accuracy. For this purpose, they are typically equipped with multiple perception sensors, including cameras, lidars, and automotive radars \cite{fan2023autonomous}. While cameras and lidars provide rich environmental information, their performance degrades significantly under adverse lighting and weather conditions \cite{fan20244d}. In contrast, millimeter-wave (mmWave) automotive radar maintains reliable sensing performance under such conditions and has therefore become a key sensing modality for automotive perception \cite{hong2021radar}. Combined with multiple-input multiple-output (MIMO) antenna technology, radar can achieve spatial resolution in compact hardware and provide object feature measurements such as range, azimuth, radial velocity, and elevation.\par

\begin{figure}[t!]
\centering
\includegraphics[width=0.45\textwidth]{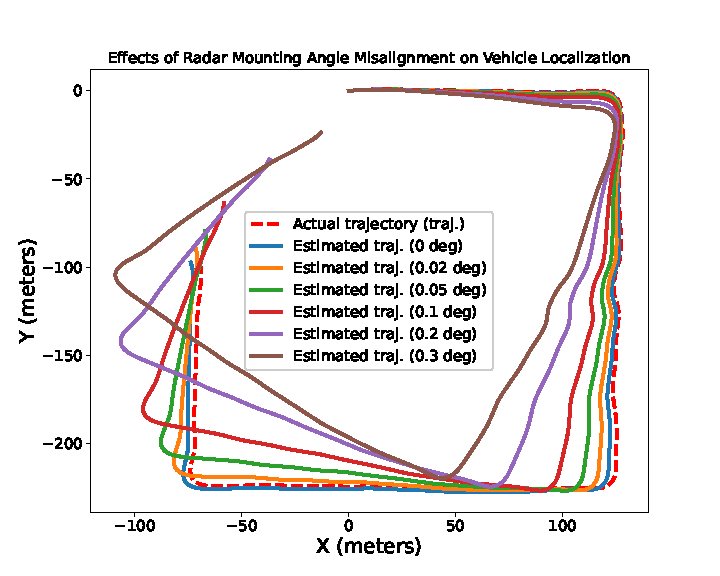}
\caption{The negative impacts of radar mounting angle misalignment on vehicle localization. The vehicle localization algorithm uses extrinsic parameters to convert the estimated radar trajectory into vehicle trajectory. However, when the extrinsic parameters are incorrect, additional errors will be injected after the conversion.}
\label{fig:chap_5_misalign_effect}
\end{figure}

Given these advantages, automotive radar has been widely adopted for tasks such as ego-motion estimation \cite{kellner2013instantaneous}, environment mapping \cite{werber2015automotive}, moving object tracking \cite{eltrass2018automotive}, and road user classification \cite{scheiner2018radar}. A common prerequisite for all these applications is accurate knowledge of the radar's extrinsic parameters, in particular its mounting angle relative to the vehicle. Although such parameters are typically set at the production stage, they can drift over time due to vibrations, collisions, or accidents. Even small deviations can severely degrade performance. For instance, as shown in Figure \ref{fig:chap_5_misalign_effect}, a misalignment of only 0.05° in radar mounting angle can cause substantial localization errors. Errors can also be amplified by radar’s long detection range, degrading crucial tasks such as mapping and sensor fusion \cite{burza2024overview}. Regular estimation of the radar mounting angle during operational driving conditions is therefore very important \cite{ yan2024sensorx2vehicle}. \par

Traditional approaches determine this angle using handheld compasses, angle sensors, or radar housings with accelerometers and actuators \cite{ham2015alignment, boyd2021automated,pinnock2024radar}. However, these methods are costly, labor-intensive, and often require skilled engineers, making them impractical for regular in-vehicle calibration. Recent research has therefore shifted toward algorithmic solutions \cite{burza2024overview}, which rely on measurements from the radar under calibration together with an additional reference sensor. While these approaches aim to enable fast and accurate mounting angle estimation that operates automatically under normal driving conditions, most existing methods have only been validated in controlled environments \cite{grebner2023self}, with special targets \cite{pervsic2017extrinsic}, or on carefully designed driving routes \cite{choi2013automatic}. \par

In practice, the key challenge in radar mounting angle estimation is not the formulation of the estimation formula itself, but the quality of the available radar measurements under operational driving conditions. In dense traffic, a large fraction of radar detections originate from moving objects unsuitable to be used as references for calibration purposes; vehicle acceleration affects Doppler estimation; and radar point clouds can be sparse and flickering. These factors collectively cause existing calibration methods to fail unless restrictive assumptions, such as static scenes or controlled routes, are imposed. \par

This work addresses this bottleneck by formulating a mounting angle estimation approach that remains applicable under normal, uncontrolled driving conditions. Based on the proposed formulation, the approach combines complementary onboard sensor measurements within a kinematic framework to enable reliable radar mounting angle estimation under challenging real-world measurement conditions. For comprehensive evaluation, this study examines two problem formulations and four estimation techniques within this framework. Experiments use the challenging \textit{RadarScenes} dataset \cite{radar_scenes_dataset}, covering more than 79 km of urban driving across varied traffic and environmental conditions, as well as varying velocity and trajectories of the ego-vehicle. To the best of our knowledge, this study demonstrates that radar mounting angles can be accurately estimated in unconstrained real-world traffic, without relying on controlled environments, dedicated radar targets, or specific driving maneuvers. \par



    
    

The rest of this paper is organized as follows. Section \ref{chap_5_related_work} reviews related work. Section \ref{chap_5_methodology} presents the proposed signal processing pipeline. Section \ref{chap_5_results} reports evaluation results and comparisons. Section \ref{chap_5_conclusions} concludes with key findings and future research directions. \par

\section{Related Works}\label{chap_5_related_work}
The main objective of radar mounting angle estimation is to determine the relative angle between the principal beam direction of the radar sensor and the thrust axis of the vehicle. To operate automatically, without human supervision, a second sensor with known extrinsic parameters is typically required as a reference. The relative angle between the radar and the reference sensor is first estimated, and then converted to the vehicle’s thrust axis using the reference sensor’s extrinsic parameters.\par

According to the type of reference sensor, existing methods can be divided into four categories: camera-based \cite{wise2023spatiotemporal,cheng2023online,scholler2019targetless}, lidar-based \cite{pervsic2021online,heng2020automatic,pervsic2017extrinsic}, radar-based \cite{cheng2023extrinsic,grebner2023self,olutomilayo2021extrinsic}, and odometry-based \cite{bobaru2022unsupervised,bao2020motion,doer2020radar}. Camera- and lidar-based approaches benefit from the high resolution and rich information provided by these optical sensors. However, their performance is strongly affected by adverse weather and lighting conditions. In addition, some require overlapping fields of view (FoV) between the radar and the reference sensor, which limits where the radar can be physically mounted \cite{scholler2019targetless,pervsic2021online}. Radar-based approaches use another automotive radar as the reference. Compared with optical sensors, they are less affected by environmental conditions. However, they often require strict conditions such as synchronized radar sensors \cite{cheng2023extrinsic,grebner2023self}, specially designed radar targets (e.g., corner reflectors) \cite{grebner20226d,olutomilayo2021extrinsic}, or overlapping FoV \cite{ikram2019automated}. These requirements are difficult to meet in real driving scenarios. In contrast, odometry-based approaches avoid these limitations. Odometry sensors operate reliably under all weather conditions and, due to their high refresh rate, do not suffer from synchronization issues. Most odometry-based methods rely on ego-velocity: by comparing the ego-velocities measured by the odometry sensor and the radar, their relative transformation can be estimated using rigid-body kinematics \cite{kellner2015joint}. This eliminates the need for challenging processing steps such as feature extraction and data association, which are common in other approaches \cite{cheng2023online,olutomilayo2021extrinsic}. Despite these advantages, odometry-based methods still face two major limitations: \par

\begin{enumerate}

    \item \textbf{Sensitivity to outliers.} To handle complex driving scenarios, most odometry-based methods use random sample consensus (RANSAC) \cite{fischler1981random} or its variants \cite{wise2021continuous} to mitigate the impact of measurement noise and moving objects (i.e., outliers) on radar motion estimation \cite{kellner2015joint,doer2020radar,bao2020motion}. However, RANSAC is an iterative algorithm with several parameters to tune and assumes that most radar measurements originate from stationary objects. Its performance degrades significantly when outliers dominate, as in dense traffic with many moving targets. Consequently, many studies evaluate their methods only in controlled environments where most surrounding objects are static \cite{izquierdo2018multi,bao2020motion,doer2020radar}, leaving performance under realistic driving scenarios less extensively validated.
    
    \item \textbf{Neglect of vehicle acceleration.} A second limitation is that most odometry-based methods ignore the effects of vehicle acceleration and deceleration \cite{bokare2017acceleration}. Acceleration causes range and Doppler migration \cite{li2024radar}, leading to inaccurate radar motion estimates \cite{zhu2025deepego+}. To mitigate this, \cite{burza2024overview} only used radar data when vehicle acceleration was below $0.5\ m/s^2$. However, such a fixed threshold is impractical in real driving scenarios.  
    
\end{enumerate}

In summary, odometry-based methods offer clear advantages over other approaches. However, the diversity of road users and the dynamic behavior of vehicles pose significant challenges for their application in real-world driving scenarios. Moreover, while different problem formulations and estimation techniques have been proposed \cite{bao2020motion,kellner2015joint}, a comprehensive comparison of their bottlenecks and trade-offs is still lacking.\par

\section{Methodology}\label{chap_5_methodology}
This section presents the proposed signal processing pipeline for radar mounting angle estimation. An overview of the proposed pipeline will first be provided. Then the design details of each processing component will be explained.\par

\subsection{Overview of Proposed Pipeline}
\begin{figure*}[!t]
\centering
\includegraphics[width=\textwidth]{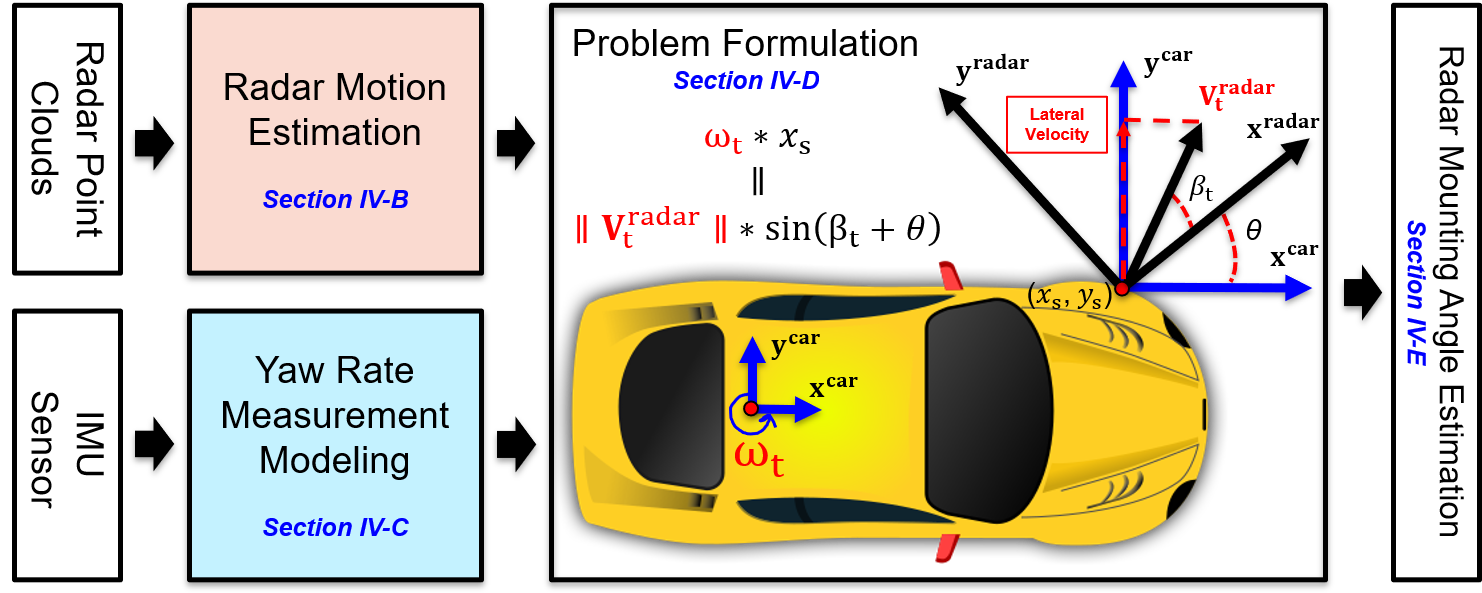}
\caption{Overview of the proposed signal processing pipeline for the problem of radar mounting angle estimation. Radar point clouds and IMU yaw-rate measurements are used as inputs, and the radar mounting angle is estimated as output. Radar motion is first estimated from the point clouds, while IMU yaw-rate measurements are modeled to account for bias and scale factor effects. The mounting angle is then obtained by enforcing a lateral velocity equality constraint within a kinematic formulation and solving a weighted least-squares problem. Here, $\mathbf{V}_t^{\mathrm{radar}}$ denotes the radar motion vector at timestamp $t$, $\omega_t$ is the vehicle yaw rate (rotational velocity), $(x_s, y_s)$ are the radar coordinates with respect to the vehicle rear center, $\beta_t$ is the direction of radar motion in the radar coordinate frame, and $\theta$ is the unknown radar mounting angle.}
\label{fig:chap_5_pipeline}
\end{figure*}

The proposed processing pipeline belongs to the category of odometry-based methods and uses an inertial measurement unit (IMU) as the reference sensor. The objective is to estimate the mounting (yaw) angle of an automotive radar system installed on a moving vehicle. The formulation is based on rigid-body kinematics and assumes normal vehicle motion without lateral side-slip, which is a standard assumption in radar odometry literature \cite{kellner2013instantaneous,schumann2021radarscenes}. As illustrated in Figure \ref{fig:chap_5_pipeline}, the pipeline takes radar point clouds and IMU yaw-rate measurements as inputs. The radar point cloud is processed by a neural network (NN) based radar motion estimator, which outputs radar ego-motion estimates together with estimated point weights \cite{zhu2023deepego}. The estimated weights are then used for variance estimation and for rejecting sparse radar frames. \par

On the IMU side, only yaw-rate measurements are used. A yaw-rate measurement model \cite{kellner2015joint} is applied to account for the IMU bias and scale factor. The yaw-rate readings are then de-biased before being used for mounting angle estimation. In the problem formulation, the proposed method exploits the velocity equality in the radar’s lateral direction. Specifically, the projection of the estimated radar motion in the lateral direction must equal the lateral velocity induced by vehicle rotation, under the no–side-slip assumption. In the final stage, radar lateral-velocity equations from multiple radar frames are combined to form an overdetermined system. The radar mounting angle and the IMU scale factor are then jointly estimated using the weighted least-squares (wLSQ) method. \par

It is important to highlight that the overall structure of the proposed pipeline is driven by the practical constrains encountered in real-world driving. Adverse weather and lighting conditions limit the reliability of optical sensors, motivating the use of an IMU as the reference sensor due to its robustness to environmental conditions, high refresh rate, and low data throughput. At the same time, radar point clouds captured in dense traffic are dominated by measurements from moving objects and are often sparse or affected by vehicle acceleration, which makes classical model-based approaches unreliable. A learning-based method is therefore employed to process radar data, handle these unfavorable conditions, and estimate radar motion. Finally, a kinematic formulation is adopted to relate radar motion and vehicle rotation without introducing vehicle-specific dynamic parameters. Together, these components form a compact pipeline required to estimate the radar mounting angle under operational driving conditions. \par


\subsection{Radar Motion Estimation}
The main objective of the radar motion estimator is to process raw radar point clouds and estimate the radar motion. In the literature, two main approaches exist for radar motion estimation: scan-matching methods \cite{rapp2015fast} and instantaneous methods \cite{kellner2013instantaneous}. It has been shown in \cite{zhu2023deepego} that, under realistic driving scenarios, instantaneous methods \cite{kellner2013instantaneous,zhu2023deepego} are generally less sensitive to point cloud sparsity and dynamic objects than scan-matching methods. As a result, most existing approaches for radar mounting angle estimation rely on instantaneous methods \cite{grebner2023self,cheng2023extrinsic,kellner2015joint,doer2020radar}. However, as discussed in Section \ref{chap_5_related_work}, these methods typically depend on random sampling (e.g., RANSAC \cite{fischler1981random} or its variants) for outlier rejection, and their performance degrades significantly when the radar point cloud contains a high proportion of outliers. In addition, RANSAC is an iterative algorithm, which can lead to poor runtime performance.  

\begin{figure*}[!t]
\centering
\includegraphics[width=\textwidth]{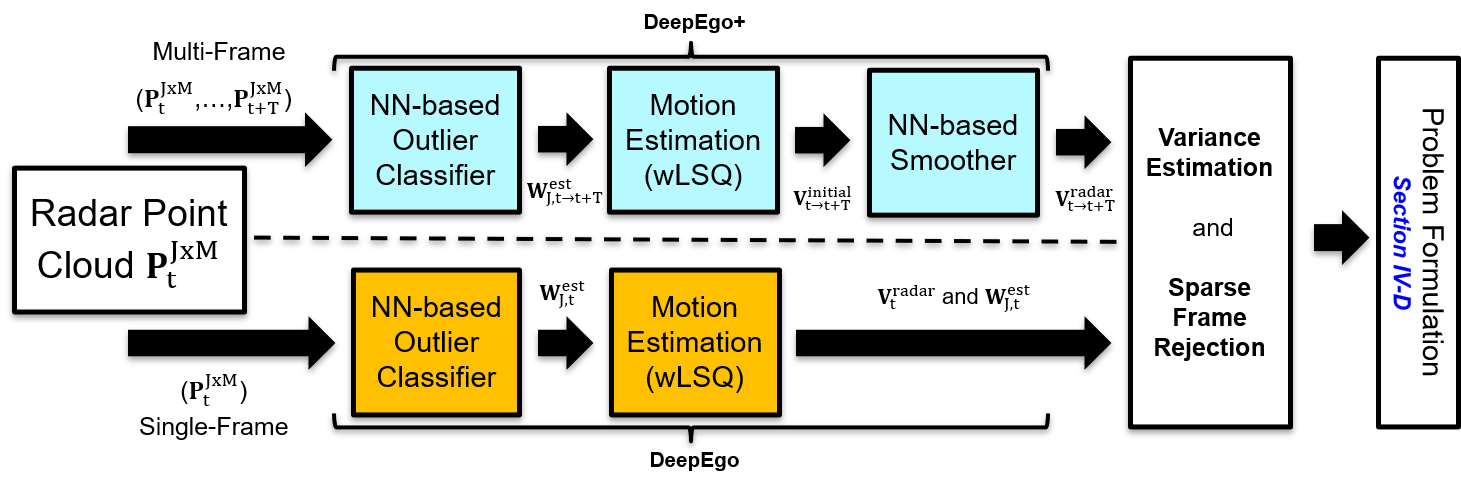}
\caption{Proposed signal processing pipeline for radar motion estimation. The estimator takes radar point clouds as input and outputs motion estimates and point weights. The weights are subsequently used to compute motion variance and reject sparse radar frames. Instead of RANSAC, two neural network-based approaches are employed: \textit{DeepEgo} \cite{zhu2023deepego} for single-frame input, and \textit{DeepEgo+} \cite{zhu2025deepego+}, which incorporates temporal layers, for multiple frames.}
\label{fig:chap_5_pipeline_rmest}
\end{figure*}

To address the above issues, the proposed pipeline employs two NN-based approaches for two scenarios: (1) a single radar frame \cite{zhu2023deepego} and (2) multiple radar frames \cite{zhu2025deepego+}. The bottom of Figure \ref{fig:chap_5_pipeline_rmest} illustrates the single-frame case. At timestamp $t$, the radar provides a point cloud that can be represented as a $J \times M$ matrix, consisting of $J$ detections (rows), each with $M$ features (e.g., range, Doppler, angle of arrival). The NN-based approach directly processes this structured input, extracts spatial sinusoidal features from the Doppler profile, and estimates a weight vector $\mathbf{W}_{J,t}^{\mathrm{est}}$ for the $J$ points. These weights are then used to eliminate outliers. The final radar motion $\mathbf{V}_t^{\mathrm{radar}}$ is estimated using the wLSQ method. \par

For the multi-frame case, motion estimates accumulated over $T$ consecutive radar frames are processed by an additional temporal NN, denoted as the ``NN-based Smoother'' in Figure \ref{fig:chap_5_pipeline_rmest}. This temporal NN captures the hidden relationship between non-zero vehicle acceleration and Doppler spectrum broadening,\footnote{Vehicle acceleration causes Doppler migration and spectrum broadening, such that the measured radial velocity of a static object no longer matches the instantaneous vehicle motion. The magnitude of the mismatch depends on the acceleration.} while also smoothing the initial estimates according to a second-order motion model. The output of the temporal NN consists of $T$ motion estimates $\mathbf{V}_{t\rightarrow t+T}^{\mathrm{radar}}$ and a weight matrix $\mathbf{W}_{J,t\rightarrow t+T}^{\mathrm{est}}$.\par

To further improve estimation reliability, the estimated radar motion and point weights are used to compute the estimation variance and to reject sparse radar frames. First, the number of inlier points $L_t$ is calculated from the weights: \par

\begin{equation}\label{chap_5_eq:1}
    \begin{split}
        L_t &= \sum_{j=1}^{J} q_{j,t}, \\
        q_{j,t} &=
        \begin{cases}
            1, & w_{j,t}^{\mathrm{est}} \geq IT, \\[6pt]
            0, & w_{j,t}^{\mathrm{est}} < IT,
        \end{cases}
    \end{split}
\end{equation}

where $IT$ denotes the inlier threshold. Using the angle $\alpha_t^l$ and Doppler $d_t^l$ measurements of the $L_t$ inliers, together with the estimated 2D radar motion $\mathbf{V}_t^{\mathrm{radar}} = \big[v_{x,t}^{\mathrm{radar}}, \; v_{y,t}^{\mathrm{radar}}\big]^\top$, the residual error $\epsilon_t$ is computed as: \par

\begin{equation}\label{chap_5_eq:2}
    \begin{split}
        \mathbf{\epsilon}_t &= \mathbf{A}_t \cdot \mathbf{V}_t^{\mathrm{radar}} - \mathbf{D}_t, \\
        \mathbf{A}_t &= \begin{bmatrix}
        \cos(\alpha_t^1) & \sin(\alpha_t^1) \\
        \vdots & \vdots \\
        \cos(\alpha_t^{L}) & \sin(\alpha_t^{L})
        \end{bmatrix}, \\
        \mathbf{D}_t &= \begin{bmatrix}
        -d_t^1 \\ \vdots \\ -d_t^{L}
        \end{bmatrix}.
    \end{split}
\end{equation}

The Doppler measurement $d_t^l$ is defined as positive when an object moves \textit{towards} the radar. Unless a sparse frame is detected, the covariance matrix of the radar motion estimate is computed as in \cite{kellner2014instantaneous}: \par

\begin{equation}\label{chap_5_eq:3}
    \mathrm{Cov}(\mathbf{V}_t^{\mathrm{radar}}) =
    \begin{cases}
        \dfrac{\mathbf{\epsilon}_t^\top \mathbf{\epsilon}_t}{L_t-2} (\mathbf{A}_t^\top \mathbf{A}_t)^{-1}, & \tfrac{L_t}{J} \geq IRT, \\[12pt]
        \mathrm{diag}(\infty, \infty), & \tfrac{L_t}{J} < IRT,
    \end{cases}
\end{equation}

where $IRT$ is a pre-determined threshold on the inlier ratio $L_t/J$. The diagonal terms of the covariance matrix, denoted $Var_{t}^{xx}$ and $Var_{t}^{yy}$, represent the variance of the estimated radar motion. These are later used in the mounting angle estimation stage to mitigate the effect of erroneous radar motion estimates and sparse radar measurements. \par

\subsection{Inertial Measurement Unit}
The proposed method uses yaw-rate readings from an IMU. Compared with other odometry sensors such as GPS, IMUs provides more reliable measurements in urban areas and in the presence of tall buildings. Nevertheless, several factors limit the accuracy of IMU yaw-rate measurements. To account for these effects, this work adopts a standard yaw-rate measurement model \cite{kellner2015joint} that includes a scale factor, a bias, and additive noise:

\begin{equation}\label{chap_5_eq:4}
    \omega_t \;=\; s\,\hat{\omega}_t \;+\; b \;+\; \nu_t,\qquad \nu_t \sim \mathcal{N}\left(0,\sigma_{\omega}^2\right),
\end{equation}

where $\omega_t$ is the measured yaw rate at time $t$, $\hat{\omega}_t$ is the true yaw rate, $s$ is the (multiplicative) scale factor, $b$ is the constant bias, and $\nu_t$ is zero-mean Gaussian noise with variance $\sigma_{\omega}^2$. Ideally, $s$ is constant over the measurement range, but it may vary with temperature, introducing systematic errors over time. Estimating $s$ is therefore important for accurate mounting-angle estimation. The bias $b$ is modeled as constant (time-invariant) in this study.

When the ego-vehicle is stationary (i.e., $\hat{\omega}_t = 0$), the measurement reduces to $\omega_t = b + \nu_t$. Over $T$ timestamps, the bias can be estimated by sample averaging:

\begin{equation}\label{chap_5_eq:5}
    \bar{b}=\frac{1}{T}\sum_{t=1}^{T}{\omega_t}
\end{equation}

\subsection{Problem Formulation}\label{chap_5_fomulation}
As illustrated in Figure \ref{fig:chap_5_pipeline}, the proposed method exploits the equality of lateral velocities at the radar position to estimate the mounting angle. Specifically,

\begin{equation}\label{chap_5_eq:6}
    \bigl\|\mathbf{V}_t^{\mathrm{radar}}\bigr\| \cdot\sin\bigl(\beta_t+\theta\bigr)
    \;=\; \frac{\tilde{\omega}_t}{s}\cdot x_s,
\end{equation}

where $\mathbf{V}_t^{\mathrm{radar}}$ is the radar motion vector at time $t$ (estimated by the NN-based motion estimator), $\beta_t$ is its direction in the radar frame, and $\theta$ is the radar mounting angle to be estimated. On the right-hand side, $\tilde{\omega}_t \triangleq \omega_t-\bar{b}$ is the debiased yaw rate, $s$ is the unknown IMU scale factor, and $x_s$ is the value of the radar's mounting location with respect to the x-axis of the origin of the vehicle. In practice, the radar mounting location $(x_s, y_s)$ is either specified by the vehicle manufacturer or can be measured with millimeter accuracy during installation \cite{radar_scenes_dataset}. Solving Equation \ref{chap_5_eq:6} for $\theta$ gives:

\begin{equation}\label{chap_5_eq:7}
    \theta \;=\; \arcsin\left(\frac{\tilde{\omega}_t\cdot x_s}{s\cdot\bigl\|\mathbf{V}_t^{\mathrm{radar}}\bigr\|}\right) \;-\; \beta_t .
\end{equation}

Assuming no scale-factor error ($s=1$), the mounting angle can be estimated over $T$ timestamps by a simple unweighted average as in \cite{kellner2015joint}:

\begin{equation}\label{chap_5_eq:8}
    \bar{\theta} \;=\; \frac{1}{T}\sum_{t=1}^{T}
    \left[
        \arcsin\left(\frac{\tilde{\omega}_t\cdot x_s}{\bigl\|\mathbf{V}_t^{\mathrm{radar}}\bigr\|}\right)
        - \beta_t
    \right].
\end{equation}

When $s\neq 1$, both $s$ and $\theta$ must be estimated jointly. Using the change of variables $s' \triangleq 1/s$, \eqref{chap_5_eq:7} can be rearranged to

\begin{equation}\label{chap_5_eq:9}
    \beta_t(\theta,s') \;=\; \arcsin\bigl(s' \cdot\chi_t\bigr)\;-\;\theta,
\end{equation}

where

\begin{equation}\label{chap_5_eq:10}
    \chi_t \;\triangleq\; \frac{\tilde{\omega}_t\cdot x_s}{\bigl\|\mathbf{V}_t^{\mathrm{radar}}\bigr\|}.
\end{equation}

Linearizing $\beta_t(\theta,s')$ at $(\theta_0,s'_0)$ via first-order Taylor expansion yields:

\begin{equation}\label{chap_5_eq:11}
\begin{aligned}
    \beta_t(\theta,s') 
    &\approx \beta_t(\theta_0,s'_0) 
      + (\theta-\theta_0)\,\frac{\partial \beta_t}{\partial \theta}(\theta_0,s'_0) \\
    &\quad + (s'-s'_0)\,\frac{\partial \beta_t}{\partial s'}(\theta_0,s'_0) \\[6pt]
    &= \arcsin\bigl(s'_0 \cdot \chi_t\bigr) - \theta 
      + \frac{(s'-s'_0)\cdot\chi_t}{\sqrt{\,1-(s'_0\cdot\chi_t)^2\,}} \, .
\end{aligned}
\end{equation}

since $\dfrac{\partial \beta_t}{\partial \theta}=-1$ and $\dfrac{\partial \beta_t}{\partial s'}=\dfrac{\chi_t}{\sqrt{1-(s'\cdot\chi_t)^2}}$.

\subsection{Mounting Angle Estimation}
From the linearized expression in Equation \ref{chap_5_eq:11}, stacking $T$ radar frames yields an overdetermined linear system ($T > 2$) in the unknowns $\theta$ (mounting angle) and $s'$ (inverse IMU scale factor). Setting the linearization point $s'_0=1$, the system can be written as:
\begin{equation}\label{chap_5_eq:12}
\begin{aligned}
    \mathbf{Y} &= \mathbf{U}\,\mathbf{X}, \\[4pt]
    \mathbf{Y} &= 
    \begin{bmatrix}
        \beta_1 - \arcsin(\chi_1) + \dfrac{\chi_1}{\sqrt{1-\chi_1^2}} \\
        \vdots \\
        \beta_T - \arcsin(\chi_T) + \dfrac{\chi_T}{\sqrt{1-\chi_T^2}}
    \end{bmatrix}, \\[10pt]
    \mathbf{U} &= 
    \begin{bmatrix}
        -1 & \dfrac{\chi_1}{\sqrt{1-\chi_1^2}} \\
        \vdots & \vdots \\
        -1 & \dfrac{\chi_T}{\sqrt{1-\chi_T^2}}
    \end{bmatrix}, \\[10pt]
    \mathbf{X} &=
    \begin{bmatrix}
        \theta \\ s'
    \end{bmatrix}.
\end{aligned}
\end{equation}

Based on the covariance of the radar motion estimates, the wLSQ solution is:

\begin{equation}\label{chap_5_eq:13}
    \begin{aligned}
        \bar{\mathbf{X}} &= \bigl(\mathbf{U}^\top \mathbf{Q}\,\mathbf{U}\bigr)^{-1}\,\mathbf{U}^\top \mathbf{Q}\,\mathbf{Y}, \\[6pt]
        \mathbf{Q} &= \mathrm{diag}\bigl(\eta_1,\dots,\eta_T\bigr), \\[6pt]
        \eta_t &= \frac{1}{Var_{t}^{xx} + Var_{t}^{yy}}.
    \end{aligned}
\end{equation}

Finally, it is worth noting that the proposed method relies on ordinary vehicle motion to enable mounting angle estimation. In particular, non-zero vehicle rotation is required so that the lateral velocity induced by vehicle yaw motion can be observed at the radar location. In addition, forward motion of the ego-vehicle is also necessary. The above conditions can be satisfied during normal driving operations, without the need of specially designed maneuvers or controlled motion patterns. \par 

\section{Results and Discussion}\label{chap_5_results}
This section presents the evaluation results of the proposed pipeline for radar mounting angle estimation. A comprehensive comparison is made with related methods from the literature, as well as with alternative problem formulations and estimation techniques. Finally, the main challenges hindering the deployment of radar mounting angle estimation algorithms in realistic driving scenarios are identified and discussed. \par

\subsection{Dataset and Evaluation Details}\label{chap_5_dataset}
\begin{figure}[t!h]
\centering
\includegraphics[width=0.42\textwidth]{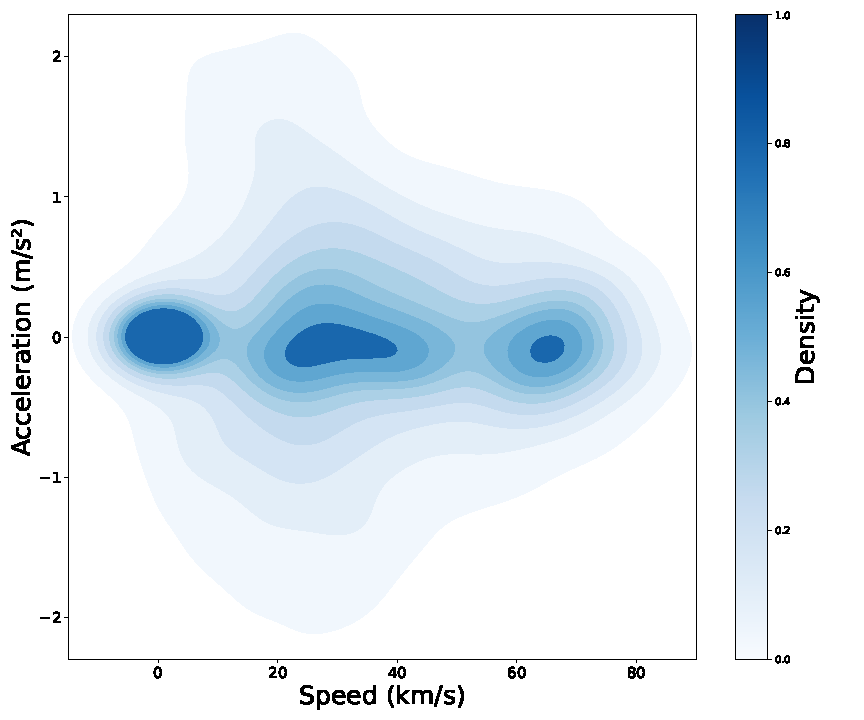}
\caption{Density distribution of ego-vehicle speed and acceleration. For visualization, only 1\% of the data (about 4.3k radar frames) was randomly sampled from the 64 selected recordings. This figure shows that the dataset used for evaluation covers a wide range of vehicle motion states during normal driving operations.}
\label{fig:chap_5_velo_acc_distribution}
\end{figure}

Unlike previous works, this study evaluates performance on the challenging real-world \textit{RadarScenes} dataset \cite{radar_scenes_dataset}. The dataset contains 158 radar recordings from four automotive radars mounted on the vehicle front, covering a variety of times and driving scenarios. Since odometry-based methods are only applicable when the ego-vehicle is moving, we use 64 recordings (as in \cite{zhu2023deepego,zhu2025deepego+}) as testing scenes. The selected recordings amount to more than 2 hours of data, corresponding to over 79 km of driving. Figure \ref{fig:chap_5_velo_acc_distribution} shows the density distribution of ego-vehicle speed and acceleration in these scenes, indicating that the dataset covers both low/high speed driving and acceleration/braking conditions. \par

In addition to challenging data, this work also conducts a comprehensive performance comparison. First, the proposed method (both single-frame, SF, and multi-frame, MF) is compared with two representative methods from the literature (Table \ref{tab:chap_5_diff_methods}). This comparison considers not only estimation accuracy, but also estimation stability and convergence speed. Moreover, this work evaluates the relative trajectory error (RTE)\footnote{RTE measures the difference between two trajectories. It first aligns the starting points of the estimated and reference trajectories, then computes the $\ell_2$ distance between their end points (unit in meters). For longer trajectories, the path can be divided into shorter segments (e.g., a 2 km trajectory divided into 50 m segments). RTE is then computed on each segment, and the final reported error (e.g., RTE\_50) is the mean over all segments.} metric \cite{zhu2023deepego}, which quantifies the impact of mounting angle misalignment on vehicle positioning and also highlights imperfections in the ground truth. Finally, this work explores alternative problem formulations and estimation techniques by modifying the proposed pipeline.  

\begin{table}[t!]
\centering
\caption{Compared methods for radar mounting angle estimation. Depending on the neural network used, the proposed method operates either on a single frame (SF) or on multiple frames (MF). wMean denotes weighted mean; wLSQ denotes weighted least squares.}
    \begin{tabular}{|c|c|c|}
    \hline
    \textbf{Method}  & \textbf{Formulation}  & \textbf{Estimation Technique}  \\ \hline
    Kellner et al. \cite{kellner2015joint}  & Lateral velocity  & RANSAC \cite{fischler1981random} + wMean  \\ \hline
    Bao et al. \cite{bao2020motion}  & Full velocity  & Kabsch algorithm \cite{lawrence2019purely}  \\ \hline
    Proposed (SF) & Lateral velocity & DeepEgo \cite{zhu2023deepego} + wLSQ \\ \hline
    Proposed (MF) & Lateral velocity & DeepEgo+ \cite{zhu2025deepego+} + wLSQ \\ \hline
    \end{tabular}
\label{tab:chap_5_diff_methods}
\end{table}

Ground-truth mounting angles are obtained from the dataset documentation \cite{radar_scenes_dataset} (referred to as `True Angle'). To ensure generalization, the 64 test scenes are excluded from model training and validation. For the proposed MF method, eight consecutive frames are accumulated and smoothed by the temporal neural network. Following the evaluation scheme in \cite{kellner2015joint}, radar data are excluded when the vehicle translational velocity is below $1\,\mathrm{m/s}$ or the rotational velocity exceeds $140^\circ/\mathrm{s}$.  

\subsection{Performance Across Diverse Driving Scenes}\label{chap_5_robustness}
\begin{figure*}[t!h]
\centering
    \subfloat[Kellner et al.: Weighted Mean \cite{kellner2015joint}]{\includegraphics[width=.5\textwidth]{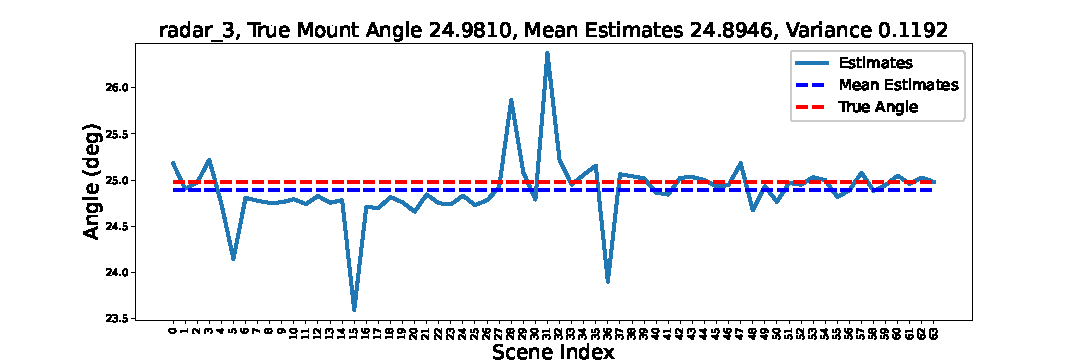}%
    \label{chap_5_robust_kellner}}
    \subfloat[Bao et al.: Kabsch Algorithm \cite{bao2020motion}]{\includegraphics[width=.5\textwidth]{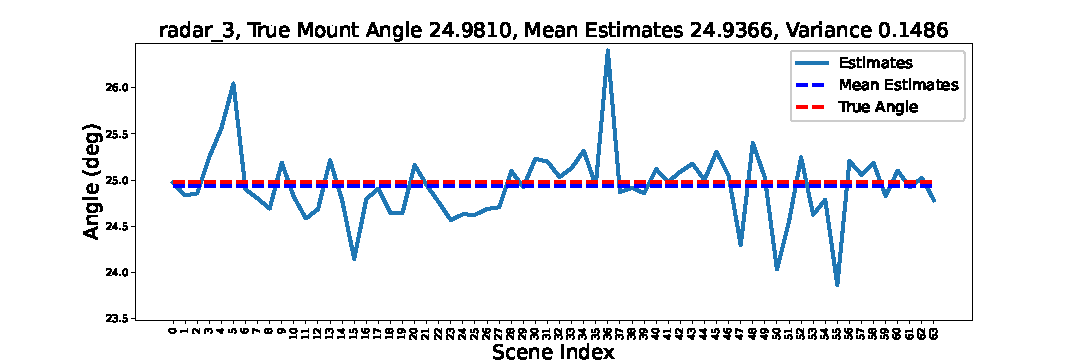}%
    \label{chap_5_robust_bao}}
    \hfill
    \subfloat[Proposed (SF)]{\includegraphics[width=.5\textwidth]{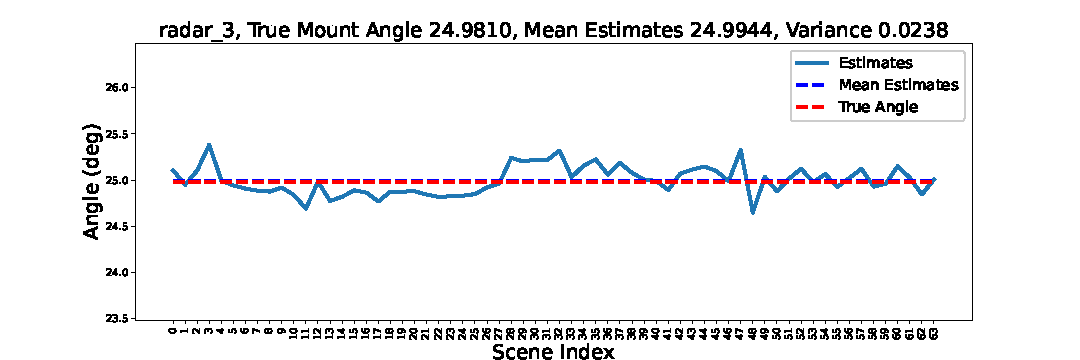}%
    \label{chap_5_robust_deepego}}
    \subfloat[Proposed (MF)]{\includegraphics[width=.5\textwidth]{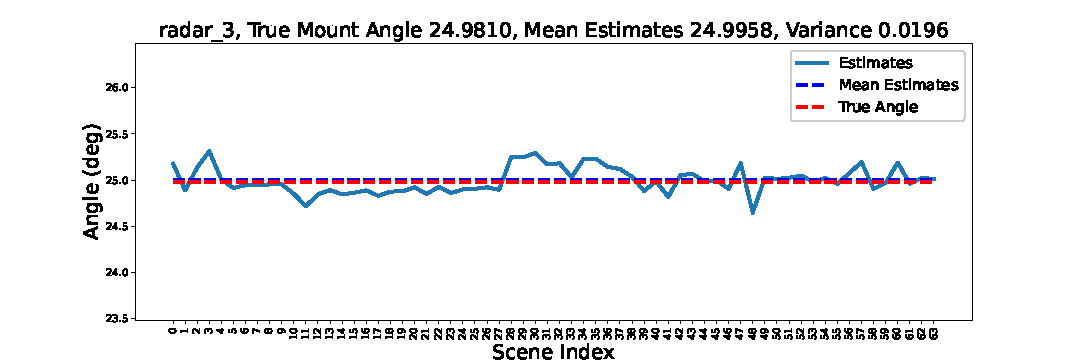}%
    \label{chap_5_robust_deepegoplus}}
\caption{Radar mounting angle estimation across 64 test scenes for different approaches. Solid blue: estimated angle per scene. Dashed blue: mean of all estimates. Dashed red: ground-truth mounting angle of Radar 3 from the \textit{RadarScenes} dataset \cite{radar_scenes_dataset}.}
\label{fig:chap_5_result_robustness}
\end{figure*}

As discussed in Section \ref{chap_5_related_work}, many previous studies evaluate radar mounting angle estimation in controlled environments (e.g., parking lots), where most surrounding objects are stationary. In contrast, real driving scenarios involve numerous moving vehicles, diverse road conditions, and a wide range of vehicle speeds and accelerations. It is therefore important to evaluate how mounting angle estimates behave across different driving scenes with varying levels of scene complexity. To this end, the estimation accuracy of all compared methods is evaluated across the 64 testing scenes. Figure \ref{fig:chap_5_result_robustness} presents the estimated mounting angle for each scene. The two baseline methods from the literature show large fluctuations around the ground-truth angle, and their performance is clearly affected by the driving environment. For example, both methods produce large errors in \textit{Scene Index 36}, which contains many moving vehicles in front of and near the ego-vehicle. Even after averaging over all 64 scenes, the mean estimates of the baselines remain far from the ground truth, with the smallest mean error still around $0.0444^\circ$. \par

In comparison, the proposed single-frame (SF) method shows substantially reduced variability across scenes, achieving an \textbf{80.0\%} reduction in error variance and a \textbf{69.8\%} improvement in mean accuracy compared with the best baseline. When multiple radar frames are available (Figure \ref{chap_5_robust_deepegoplus}), the proposed multi-frame (MF) method further exploits temporal motion correlations to mitigate the effects of outliers and non-zero vehicle acceleration, resulting in more consistent estimates across scenes. It should be noted that all results in this section are based on data from a forward-looking radar (`Radar 3' in the dataset), which is particularly challenging because of the large number of moving objects in its FoV. Consequently, even with the proposed method, the estimates cannot be made fully consistent across all scenes, and the variance cannot be driven to zero. Preliminary experiments suggest that fusing data from multiple radars could alleviate this limitation, but a full exploration of multi-radar fusion is left for future work. \par   

\subsection{Estimation Accuracy}\label{chap_5_accuracy}
Unlike mounting location, the perception performance of automotive radar is highly sensitive to small misalignments of the mounting angle, since modern radar systems can detect objects at long range. Consequently, the proposed method must provide accurate angle estimates so that downstream radar-based tasks are not degraded. In addition, because modern vehicles are typically equipped with multiple radars mounted at different positions and orientations, the proposed method should perform consistently across sensors. Table \ref{tab:chap_5_accuracy} reports the estimation results over 64 test scenes and four radar sensors. At first glance, the mean and variance of the estimation error are much higher for Radar 2 and Radar 3 compared to Radar 1 and Radar 4. This is expected, since Radars 2 and 3 are forward-looking and therefore observe more dynamic objects. Nevertheless, both proposed methods outperform the two baselines from the literature.  

The proposed single-frame approach achieves substantial accuracy improvements, while the multi-frame approach further reduces variance by smoothing motion estimates across frames. Interestingly, when compared with the dataset-provided `True Angle', the SF and MF methods show similar mean accuracy, even though the MF method would theoretically be expected to perform better. A plausible explanation is that the `True Angle' values in the dataset \cite{radar_scenes_dataset} are quantized with limited resolution. Unfortunately, the actual resolution is not documented. Section \ref{chap_5_gt} provides an alternative evaluation to indirectly assess how close the proposed estimates are to the actual mounting angles.  \par

\begin{table*}[t!h]
\caption{Performance comparison over 64 testing scenes and 4 radars. Mean and variance of the estimated mounting angle are reported. $|\Delta \text{Min}|$ is the absolute difference between the `True Angle' and the best mean estimate among the four methods. $|\Delta \text{Max}|$ is the absolute difference for the worst case. Blue highlights the best-performing values, while red highlights the worst.}
\centering
\begin{tabular}{|c|cc|cc|cc|cc|}
\hline
\multirow{2}{*}{\textbf{Methods}} & \multicolumn{2}{c|}{\textbf{Radar 1}}       & \multicolumn{2}{c|}{\textbf{Radar 2}}       & \multicolumn{2}{c|}{\textbf{Radar 3}}       & \multicolumn{2}{c|}{\textbf{Radar 4}}       \\ \cline{2-9} 
                     & \multicolumn{1}{c|}{Mean}  & Variance  & \multicolumn{1}{c|}{Mean}  & Variance  & \multicolumn{1}{c|}{Mean}  & Variance & \multicolumn{1}{c|}{Mean} & Variance \\ \hline
Weighted Mean \cite{kellner2015joint}                   & \multicolumn{1}{c|}{\textcolor{red}{\textbf{-85.1107}}\textdegree} & 0.0051 & \multicolumn{1}{c|}{-24.8973\textdegree} & 0.0305 & \multicolumn{1}{c|}{\textcolor{red}{\textbf{24.8946}}\textdegree} & 0.1192 & \multicolumn{1}{c|}{85.0030\textdegree} & 0.0059 \\ \hline
Kabsch Algorithm \cite{bao2020motion}                  & \multicolumn{1}{c|}{-85.1102\textdegree} & \textcolor{red}{\textbf{0.0290}} & \multicolumn{1}{c|}{\textcolor{red}{\textbf{-24.7806}}\textdegree} & \textcolor{red}{\textbf{0.8037}} & \multicolumn{1}{c|}{24.9366\textdegree} & \textcolor{red}{\textbf{0.1486}} & \multicolumn{1}{c|}{\textcolor{red}{\textbf{84.9930}}\textdegree} & \textcolor{red}{\textbf{0.0442}} \\ \hline
Proposed (SF)                  & \multicolumn{1}{c|}{\textcolor{blue}{\textbf{-85.0418}\textdegree}} & 0.0035 & \multicolumn{1}{c|}{-25.0012\textdegree} & 0.0286 & \multicolumn{1}{c|}{\textcolor{blue}{\textbf{24.9944}\textdegree}} & 0.0238 & \multicolumn{1}{c|}{\textcolor{blue}{\textbf{85.0256}}\textdegree} & 0.0027 \\ \hline
Proposed (MF)                  & \multicolumn{1}{c|}{-85.0467\textdegree} & \textcolor{blue}{\textbf{0.0025}} & \multicolumn{1}{c|}{\textcolor{blue}{\textbf{-24.9988}\textdegree}} & \textcolor{blue}{\textbf{0.0184}} & \multicolumn{1}{c|}{24.9958\textdegree} & \textcolor{blue}{\textbf{0.0196}} & \multicolumn{1}{c|}{85.0286\textdegree} & \textcolor{blue}{\textbf{0.0021}} \\ \hline
True Angle                   & \multicolumn{2}{c|}{-85.0376\textdegree}      & \multicolumn{2}{c|}{-24.9916\textdegree}      & \multicolumn{2}{c|}{24.9810\textdegree}      & \multicolumn{2}{c|}{85.0269\textdegree}      \\ \hline
$|\Delta \text{Min}|$ in Angle               & \multicolumn{2}{c|}{\textcolor{blue}{\textbf{0.0042\textdegree}}}      & \multicolumn{2}{c|}{\textcolor{blue}{\textbf{0.0072\textdegree}}}      & \multicolumn{2}{c|}{\textcolor{blue}{\textbf{0.0134\textdegree}}}      & \multicolumn{2}{c|}{\textcolor{blue}{\textbf{0.0013\textdegree}}}      \\ \hline
$|\Delta \text{Max}|$ in Angle               & \multicolumn{2}{c|}{\textcolor{red}{\textbf{0.0731\textdegree}}}      & \multicolumn{2}{c|}{\textcolor{red}{\textbf{0.2110\textdegree}}}      & \multicolumn{2}{c|}{\textcolor{red}{\textbf{0.0864\textdegree}}}      & \multicolumn{2}{c|}{\textcolor{red}{\textbf{0.0339\textdegree}}}      \\ \hline
\end{tabular}
\label{tab:chap_5_accuracy}
\end{table*}

\subsection{Convergence}\label{chap_5_speed}
The previous results demonstrated that the proposed method achieves accurate radar mounting angle estimation with low scene-to-scene variability. However, those results were either evaluated per scene (Figure \ref{fig:chap_5_result_robustness}) or averaged across all scenes (Table \ref{tab:chap_5_accuracy}). For realistic driving scenarios, it is equally important to assess how quickly (in terms of seconds) the estimation converges to a stable value.

\begin{figure*}[t!h]
    \centering
    \setkeys{Gin}{width=0.45\textwidth}
    \subfloat[MAE with different time intervals
    \label{fig:chap_5_converge_mean}]{\includegraphics{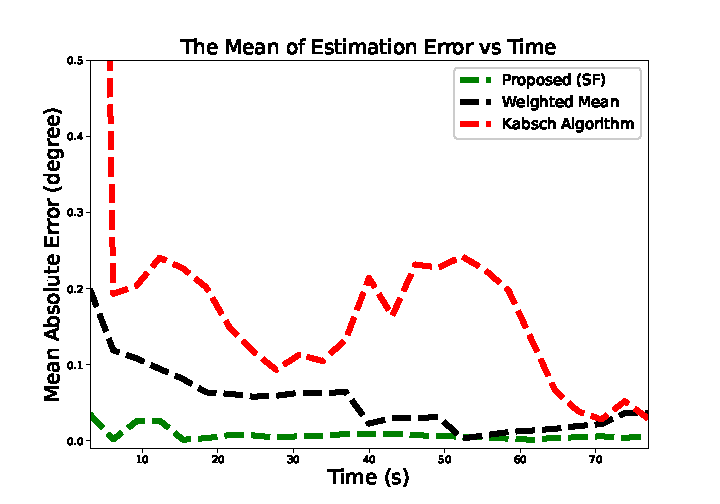}}
    \hfill
    \subfloat[Variance with different time intervals
    \label{fig:chap_5_converge_var}]{\includegraphics{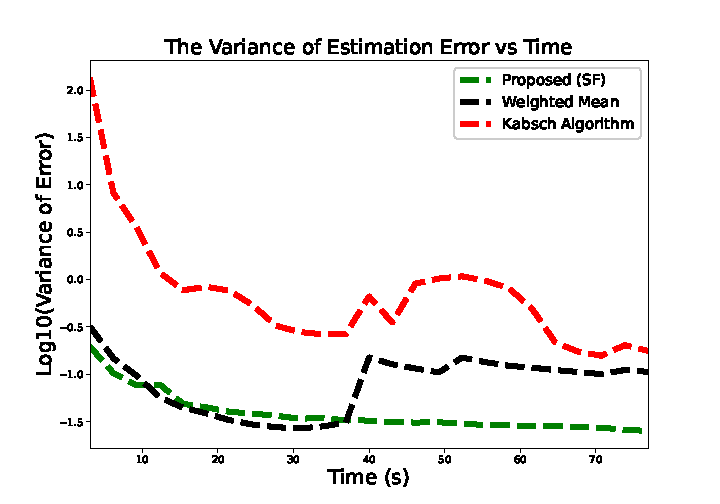}}
    \caption{Convergence behavior of different methods with respect to the number of radar frames. (a) Mean absolute error (MAE). (b) Error variance. The red dashed line denotes the Kabsch approach \cite{bao2020motion}, the black dashed line denotes the weighted mean method \cite{kellner2015joint}, and the solid green line denotes the proposed method. Each test scene is divided into shorter segments of different lengths; mounting angles are estimated for each segment, and MAE and variance are then computed.}
    \label{fig:chap_5_converge_acc_var}
\end{figure*}

To examine this, Figure \ref{fig:chap_5_converge_acc_var} presents the mean absolute error (MAE) and error variance as functions of the time interval, i.e., the number of frames processed by each method. For MAE (Figure \ref{fig:chap_5_converge_mean}), the proposed method achieves low estimation error within only a few seconds of driving, whereas the two reference methods require much more time before producing comparable accuracy. For error variance (Figure \ref{fig:chap_5_converge_var}), the proposed method consistently benefits from longer time intervals, yielding smaller variances. In contrast, the baseline methods show no clear convergence: their variance decreases initially, but then increases and fluctuates strongly with longer intervals.  


Finally, it is worth noting that these results are based on the 64 test scenes from Radar 3 of the \textit{RadarScenes} dataset, the forward-looking radar with many moving objects in its field of view. The result demonstrates that the proposed method can produce accurate mounting angle estimates with low variance within a short period of driving, even under challenging traffic conditions and with variation in velocity and trajectories of the ego-vehicle. \par

\subsection{Trajectory Error}\label{chap_5_gt}
The vehicle trajectory can be reconstructed using recorded timestamps, estimated radar motion, and radar extrinsic parameters (i.e., mounting angle and position). Intuitively, the worse the mounting angle estimate, the greater the deviation of the reconstructed trajectory from the true trajectory. Thus, trajectory error provides an indirect measure of how close an estimated mounting angle is to the actual angle. This comparison can also include the dataset-provided mounting angle (previously referred to as the `True Angle'), although its resolution is undocumented. For quantitative evaluation, this section adopts the RTE metric, which measures the discrepancy between the estimated vehicle trajectory and the ground-truth trajectory provided by the on-vehicle odometry system. Because both radar motion estimates and mounting angles influence trajectory reconstruction, in this experiment the estimated radar motion is fixed (controlled variable), while the radar mounting angle is varied (dependent variable). \par

Results are reported in Table \ref{tab:chap_5_rte}. As expected, larger angle estimation errors lead to larger RTE values. For example, Radar 1 and Radar 3 with the weighted-mean method \cite{kellner2015joint} exhibit higher angle errors in Table \ref{tab:chap_5_accuracy}, which correspond to higher RTE values here. More importantly, the proposed methods consistently yield lower RTE than the baselines, demonstrating that improved angle estimates directly enhance trajectory accuracy. Interestingly, the proposed methods also achieve slightly lower RTE than the dataset-provided `True Angle'. While this difference is small, it suggests that the proposed approach may in fact yield more accurate angles than those documented in the dataset. Nonetheless, as emphasized in \cite{kellner2015joint}, determining mounting angles with high precision is inherently difficult. The RTE metric therefore provides a useful indirect validation of the proposed method’s accuracy. \par 

\begin{table}[t!h]
\caption{Relative Trajectory Error (RTE) in meters \cite{zhu2023deepego}, averaged over 64 testing scenes. RTE measures the discrepancy between the estimated and ground-truth vehicle trajectories. The estimated trajectory is computed from timestamps, radar motions, and mounting angles. Radar motion is fixed (controlled), while mounting angle is varied (dependent). The baseline RTE is computed from the dataset-provided mounting angles \cite{radar_scenes_dataset}. Blue highlights the best-performing values, while red highlights the worst.}
\centering
\begin{tabular}{|c|cl|cl|cl|cl|}
\hline
\textbf{Sources of angle} & \multicolumn{2}{c|}{\textbf{Radar 1}}  & \multicolumn{2}{c|}{\textbf{Radar 2}}  & \multicolumn{2}{c|}{\textbf{Radar 3}}  & \multicolumn{2}{c|}{\textbf{Radar 4}}  \\ \hline
Weighted Mean \cite{kellner2015joint}  & \multicolumn{2}{c|}{\textcolor{red}{\textbf{11.98}}} & \multicolumn{2}{c|}{13.41} & \multicolumn{2}{c|}{\textcolor{red}{\textbf{12.03}}} & \multicolumn{2}{c|}{6.31} \\ \hline
Kabsch Algorithm \cite{bao2020motion}  & \multicolumn{2}{c|}{11.92} & \multicolumn{2}{c|}{\textcolor{red}{\textbf{26.98}}} & \multicolumn{2}{c|}{7.50} & \multicolumn{2}{c|}{\textcolor{red}{\textbf{6.81}}} \\ \hline
Proposed (SF)  & \multicolumn{2}{c|}{7.30} & \multicolumn{2}{c|}{5.96} & \multicolumn{2}{c|}{6.45} & \multicolumn{2}{c|}{\textcolor{blue}{\textbf{6.24}}} \\ \hline
Proposed (MF)  & \multicolumn{2}{c|}{\textcolor{blue}{\textbf{7.28}}} & \multicolumn{2}{c|}{\textcolor{blue}{\textbf{5.95}}} & \multicolumn{2}{c|}{6.51} & \multicolumn{2}{c|}{6.32} \\ \hline
Baseline RTE  & \multicolumn{2}{c|}{7.37} & \multicolumn{2}{c|}{6.05} & \multicolumn{2}{c|}{\textcolor{blue}{\textbf{5.99}}} & \multicolumn{2}{c|}{6.27} \\ \hline
\end{tabular}
\label{tab:chap_5_rte}
\end{table}

\subsection{Further Exploration}\label{chap_5_explore}
As detailed in Section \ref{chap_5_methodology}, the proposed method estimates the radar mounting angle by enforcing the equality of lateral velocities at the radar position. The formulation also incorporates the IMU measurement model, enabling joint estimation of the IMU scale factor and the mounting angle. However, several alternative approaches exist. For example, if both an IMU sensor and a DGPS system are available, then a full-velocity model can be constructed \cite{bao2020motion} and the mounting angle estimated using the Kabsch algorithm. If the IMU scale factor is close to 1, the IMU model can be ignored and the problem simplified to an averaging scheme \cite{kellner2015joint}. Moreover, the orthogonal distance regression (ODR) can be applied to extend least squares (LSQ) to cases where errors also exist in the independent variables.  \par

To better understand the performance trade-offs, the proposed method was modified accordingly while still using the frame weights from Equation (\ref{chap_5_eq:13}). Results are shown in Table \ref{tab:chap_5_diff_techniques}. Compared with Table \ref{tab:chap_5_accuracy}, the performance gap between different formulations is now much smaller. Since the main difference between these methods and the baselines from the literature lies in the radar motion and frame weighting, this suggests that a major limiting factor in practical mounting angle estimation is the high proportion of outliers and sparse radar measurements. For the chosen dataset, the weighted mean solution appears most practical: it uses the simplest model, runs the fastest, and provides performance comparable to more complex techniques. If the IMU scale factor deviates significantly from 1, the weighted LSQ should instead be used. Finally, if the radar has limited resolution in azimuth and radial velocity, weighted ODR may yield better results. \par

\begin{table}[t!h]
\caption{Performance comparison of the proposed method (MF version) with different problem formulations and estimation techniques. For each radar, results are averaged over 64 testing scenes. $|\Delta \text{Min}|$ is the absolute difference between the `True Angle' and the best estimate out of the compared methods. Blue highlights the best-performing values, while red highlights the worst.}
\centering
    \begin{tabular}{|c|c|c|c|c|}
    \hline
    \textbf{Methods}  & \textbf{Radar 1}  & \textbf{Radar 2}  & \textbf{Radar 3}  & \textbf{Radar 4 } \\ \hline
    Weighted LSQ  & \textcolor{blue}{\textbf{-85.0467}}\textdegree  & \textcolor{blue}{\textbf{-24.9988}}\textdegree  & \textcolor{blue}{\textbf{24.9958}}\textdegree  & 85.0286\textdegree \\ \hline
    Weighted Kabsch & -85.0493\textdegree & -25.0027\textdegree & 25.0071\textdegree & 85.0263\textdegree \\ \hline
    Weighted Mean & -85.0476\textdegree & -25.0005\textdegree & 25.0001\textdegree & \textcolor{blue}{\textbf{85.0269}}\textdegree \\ \hline
    Weighted ODR & -85.0469\textdegree & -25.0107\textdegree & 25.0043\textdegree & 85.0285\textdegree \\ \hline
    True Angle & -85.0376\textdegree & -24.9916\textdegree & 24.9810\textdegree & 85.0269\textdegree \\ \hline
    $|\Delta \text{Min}|$ & \textcolor{blue}{\textbf{0.0091\textdegree}} & \textcolor{blue}{\textbf{0.0072\textdegree}} & \textcolor{blue}{\textbf{0.0148\textdegree}} & \textcolor{blue}{\textbf{0.0000\textdegree}} \\ \hline
    \end{tabular}
    \label{tab:chap_5_diff_techniques}
\end{table}

\section{Conclusions}\label{chap_5_conclusions}
This paper presented a novel signal processing pipeline to address the problem of estimating radar mounting angles under operational driving conditions. Accurate external calibration of automotive radars, and in particular their mounting angles, is crucial for the safe operation of autonomous vehicles. To address this problem, an odometry-based approach was proposed that combines a neural network-based radar motion estimator with an IMU measurement model for bias and scale factor compensation. The mounting angle and IMU scale factor are then jointly estimated using a wLSQ formulation based on a Taylor-series linearization. The proposed pipeline was validated on the challenging \textit{RadarScenes} dataset, which includes diverse traffic scenes as well as velocity and trajectory variations of the ego-vehicle.

Experimental results demonstrate that the method achieves accurate mounting angle estimates with low variability across diverse and realistic driving scenarios, while avoiding the need for controlled environments, specially designed radar targets, or tailored driving routes. In addition, the estimation converges within a short period of driving time. Although the formulation involves a single linearization step, experimental results show that this is sufficient in practice, requiring no further iteration. For future work, extending the framework to sensor fusion represents a promising direction. Combining multiple radar sensors or integrating complementary modalities (e.g., cameras or lidars) may further mitigate the impact of dynamic objects in the field of view and enhance calibration accuracy under complex real-world conditions.\par

\bibliographystyle{IEEEtran}
\bibliography{reference} 
\vspace{-3 mm}
\section*{Biography Section}
\vspace{-13 mm}
\begin{IEEEbiography}[{\includegraphics[width=1in,height=1.25in,clip,keepaspectratio]{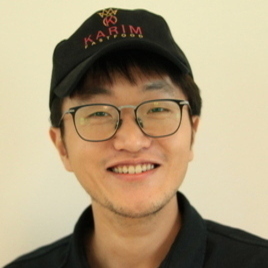}}]{Simin Zhu}
received his BSc degree in Electrical Engineering and Automation from the Central South University in 2016. Afterward, he worked for 1.5 years as a hardware engineer at Huawei Technology Co. Ltd. In 2019, Simin started his master's study at Delft University of Technology (TU Delft). During his master's program, he specialized in radar signal processing and machine learning. In November of 2021, he completed his master's thesis and graduated from the Microwave Sensing, Signals and Systems (MS3) group at TU Delft. In December 2021, he continued his research in the MS3 group as a Ph.D. candidate.
\end{IEEEbiography}\vspace{-11 mm}
\begin{IEEEbiography}[{\includegraphics[width=1in,height=1.25in,clip,keepaspectratio]{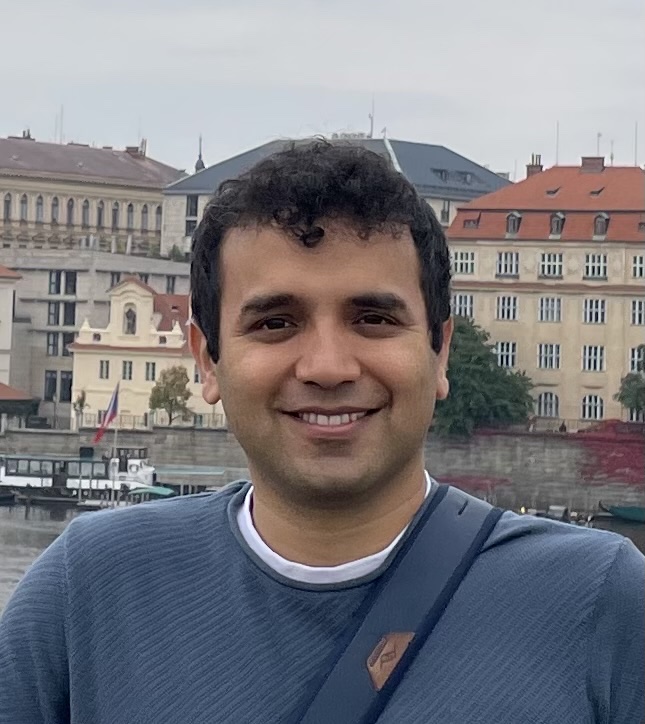}}]{Satish Ravindran}
has around 12 years of experience developing AI solutions for different industries such as Autonomous Driving, Intelligent Traffic Sensing (ITS) and IoT. He has worked on a wide spectrum of applications in AI including NLP, Computer Vision and Radar Processing. He joined NXP in 2018 and is currently the AI Technical Lead for Radar Innovations working in the NXP R\&D division. He has led the development of a comprehensive portfolio of AI applications at all stages of the radar processing chain, from signal processing to perception. He is also helping in the definition of the next generation of NXP SoCs and has multiple patents and papers published in radar signal processing and AI solutions.
\end{IEEEbiography}\vspace{-11 mm}
\begin{IEEEbiography}[{\includegraphics[width=1in,height=1.25in,clip,keepaspectratio]{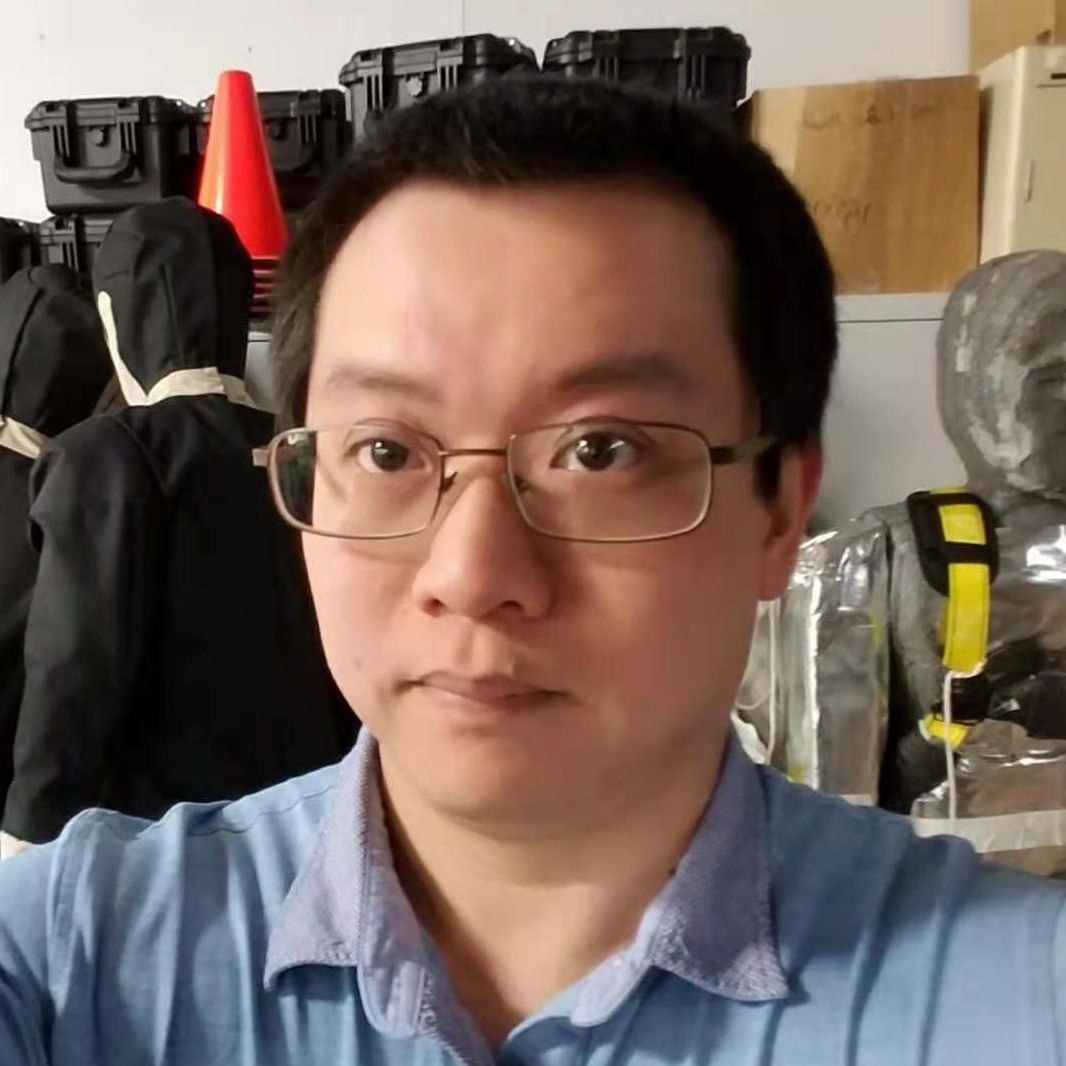}}]{Lihui Chen}
Lihui Chen received his BSc degree in Applied Mathematics from Sun Yat-sen University, China, in 1999. And PhD degree in Electrical and Electronic Engineering from Loughborough University, UK, in 2006. He has experience in machine learning for 18 years, with 12 years focusing on research of automotive ADAS products, including surround vision, driver monitoring, and HD mapping. He authored and co-authored 13 patents, with 5 systems delivered for production. He joined NXP's R\&D division in 2022 as the radar perception project lead. His research interest include AI solutions for perceptions with different sensors, and neutral network optimization for edge devices.
\end{IEEEbiography}\vspace{-11 mm}
\begin{IEEEbiography}[{\includegraphics[width=1in,height=1.25in,clip,keepaspectratio]{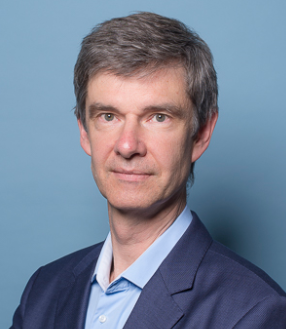}}]{Alexander G. Yarovoy}
(FIEEE’ 2015) graduated from the Kharkov State University, Ukraine, in 1984 with the Diploma with honor in radiophysics and electronics. He received the Candidate Phys. \& Math. Sci. and Doctor Phys. \& Math. Sci. degrees in radiophysics from the same university in 1987 and 1994, respectively.
In 1987 he joined the Department of Radiophysics at the Kharkov State University as a Researcher and became a Full Professor there in 1997. From September 1994 through 1996 he was with Technical University of Ilmenau, Germany as a Visiting Researcher. Since 1999 he is with the Delft University of Technology, the Netherlands. Since 2009 he leads there a chair of Microwave Sensing, Systems and Signals.
His main research interests are in high-resolution radar, microwave imaging and applied electromagnetics (in particular, UWB antennas). He has authored and co-authored more than 600 scientific or technical papers, eleven patents and fourteen book chapters. He is the recipient of the European Microwave Week Radar Award for the paper that best advances the state-of-the-art in radar technology in 2001 (together with L.P. Ligthart and P. van Genderen) and in 2012 (together with T. Savelyev). In 2023 together with Dr. I.Ullmann, N. Kruse, R. Gündel and Dr. F. Fioranelli he got the best paper award at IEEE Sensor Conference. In 2010 together with D. Caratelli Prof. Yarovoy got the best paper award of the Applied Computational Electromagnetic Society (ACES).
In the period 2008-2017 Prof. Yarovoy served as Director of the European Microwave Association (EuMA). He is and has been serving on various editorial boards such as that of the IEEE Transaction on Radar Systems. From 2011 till 2018 he served as an Associated Editor of the International Journal of Microwave and Wireless Technologies. He has been member of numerous conference steering and technical program committees. He served as the General TPC chair of the 2020 European Microwave Week (EuMW’20), as the Chair and TPC chair of the 5th European Radar Conference (EuRAD’08), as well as the Secretary of the 1st European Radar Conference (EuRAD’04). He served also as the co-chair and TPC chair of the Xth International Conference on GPR (GPR2004).
\end{IEEEbiography}\vspace{-11 mm}
\begin{IEEEbiography}[{\includegraphics[width=1in,height=1.25in,clip,keepaspectratio]{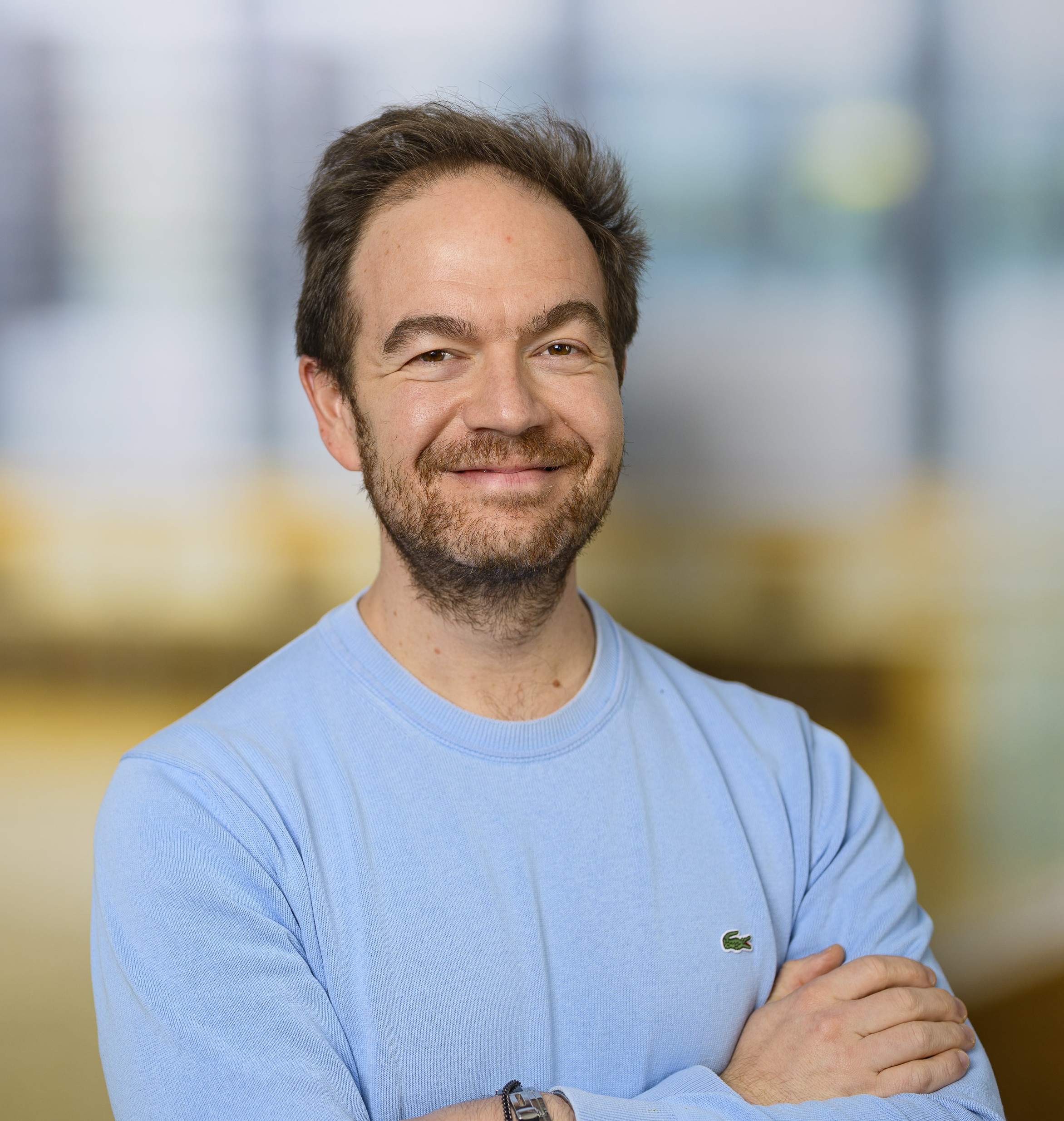}}]{Francesco Fioranelli}
(M'15–SM'19) received the Ph.D. degree with Durham University, Durham, UK, in 2014. He is currently an Associate Professor at TU Delft, The Netherlands, and was an Assistant Professor with the University of Glasgow (2016–2019), and a Research Associate at University College London (2014–2016). 

His research interests include the development of radar systems and automatic classification for human signatures analysis in healthcare and security, drones and UAVs detection and classification, and automotive radar. He has authored over 190 peer-reviewed publications, edited the books on “Micro-Doppler Radar and Its Applications” and "Radar Countermeasures for Unmanned Aerial Vehicles" published by IET-Scitech in 2020, received four best paper awards and the IEEE AESS Fred Nathanson Memorial Radar Award 2024.
\end{IEEEbiography}
\vfill
\end{document}